\documentclass[12pt,preprint]{aastex}
\usepackage{amssymb}
\usepackage{amsmath}
\usepackage{graphicx}
\usepackage{subfigure}
\usepackage{wrapfig}
\usepackage[colorlinks,linkcolor=blue,anchorcolor=green,citecolor=blue]{hyperref}

\begin{document}

\def\unit#1{\mathord{\thinspace\rm #1}}

\title{PTF11agg as the First Evidence for Reverse Shock Emission from a
Postmerger Millisecond Magnetar}
\author{Ling-Jun Wang, and Zi-Gao Dai}

\begin{abstract}
Based on the stiff equations of state of neutron stars (NS) and the
discovery of high-mass NSs, a NS-NS merger will leave behind, with high
probabilities, a rapidly rotating massive magnetar. The central magnetar
will dissipate its rotational energy to the outflow by injecting Poynting
flux, which will become lepton-dominated so that a long-lasting reverse
shock (RS) is developed. We calculate the emission of the RS as well as the
emission of forward shock (FS) and find that in most cases the RS emission
is stronger than FS emission. It is found that the recently discovered
transient, PTF11agg, can be neatly accounted for by the RS emission powered
by a millisecond magnetar. Other alternative models have been considered and
cannot explain the observed light curves well. We therefore suggest that
PTF11agg be the first evidence for RS emission from a postmerger millisecond
magnetar.
\end{abstract}



\affil{School of Astronomy and Space Science, Nanjing University, Nanjing,
China; dzg@nju.edu.cn}

\affil{Key laboratory of Modern Astronomy and Astrophysics (Nanjing
University), Ministry of Education, Nanjing 210093, China}

\section{Introduction}

\label{sec-intro}

When binary neutron stars (BNS) merge, a black hole is usually assumed to be
formed \citep{rev-bin}. With the theoretical work on stiff equations of
state and the discovery of massive neutron stars \citep{lattimer12}, several
authors \citep{dai06,zhang13} suggest that a stable massive neutron star
(NS) may be formed as a post-merger product. This suggestion is supported by
numerical-relativity simulations \citep{hotokezaka13,magnetar-sim}. Because
the newborn NSs are differentially rotating rapidly, the onset of
magneto-rotational instability could boost the magnetic field of such NSs to
magnetar levels \citep{duncan92,kluzniak98,dai98a}. Energy injection from
millisecond magnetars is also invoked to account for the unusual X-ray
emission following some short gamma-ray bursts (SGRBs) %
\citep{dai06,fan06,rowlinson10,rowlinson13}.

The electromagnetic signatures of NS-NS mergers include SGRBs %
\citep{eichler89,barthelmy05,gehrels05,rezzolla11}, radio afterglows %
\citep{nakar11,metzger12,rosswog13,piran13}, day-long optical macronovae %
\citep{li98,kulkarni05,rosswog05,metzger10,roberts11,metzger12}, and
possible X-ray emissions \citep{palenzuela13} due to the interaction of the
NS magnetospheres during the inspiral and merger. \cite{zhang13} recently
suggested that, by the formation of rapidly spinning magnetars, there is a
significant fraction of NS-NS mergers that may be detected as bright X-ray
transients associated with gravitational wave bursts (GWBs) without apparent
association of SGRBs. Subsequently, based on the energy injection scenario
proposed by \cite{dai98b}, \cite{gao13} considered the rich electromagnetic
signatures of the forward shock (FS) driven by ejecta subject to continuous
injection of Poynting flux from the central magnetars.

We here consider the electromagnetic signatures not only of FS that was
considered by \cite{gao13}, but also of reverse shock (RS) because it is
more likely that the magnetar wind would be dominated by ultra-relativistic
leptons ($e^{+}e^{-}$ pairs) within radius $\sim 10^{17}\unit{cm}$ %
\citep{dai04,coroniti90,michel94}.

The structure of this paper is as follows. In Section \ref{sec-model} we
outline our model, analytical method, and results. Section \ref{sec-PTF11agg}
presents the numerical method and its application to the recently discovered
transient source PTF11agg. In Section \ref{sec-discuss} we discuss other
alternative models to interpret PTF11agg.

\section{The Model}

\label{sec-model}

The basic picture of the model is illustrated in Figure 1 of \cite{gao13}.
The merger of BNSs ejects a mildly anisotropic outflow with typical velocity
$0.1-0.3c$ and mass $M_{\mathrm{ej}}\sim 10^{-4}-10^{-2}M_{\odot }$ %
\citep{rezzolla10,hotokezaka13,rosswog13}. The onset of Poynting flux
launched $\sim 10\unit{s}$ later catches up the ejecta and crosses the
ejecta in $t_{\Delta }\sim 3\unit{s}L_{0,47}^{-1/2}\Delta _{7}^{1/2}M_{%
\mathrm{ej},-3}^{1/2}$ \citep{gao13}, where $L_{0}\equiv \xi L_{\mathrm{sd}%
,0}$ is the power injected into the ejecta by the magnetar wind, $L_{\mathrm{%
sd},0}=1\times 10^{47}\unit{erg}\unit{s}%
^{-1}P_{0,-3}^{-4}B_{p,14}^{2}R_{6}^{6}$ is the luminosity of the central
magnetar. Here we adopt the usual convention $Q=10^{n}Q_{n}$. The value of $%
\Delta \sim 10^{7}\unit{cm}$ is just a reasonable suggestion and is
unimportant here because our results do not depend on this value. It is
known that the Poynting flux from the magnetar via magnetic dipole radiation
is nearly isotropic. Thus, the Poynting flux will always catch up with the
ejecta, though the latter could be asymmetric.

The dynamics of the blast wave is determined by%
\begin{equation}
L_{0}\min \left( t,T_{\mathrm{sd}}\right) =\left( \gamma -\gamma _{\mathrm{ej%
},0}\right) M_{\mathrm{ej}}c^{2}+2\left( \gamma ^{2}-1\right) M_{\mathrm{sw}%
}c^{2},  \label{dynamics}
\end{equation}%
where $T_{\mathrm{sd}}$ is the spin-down time of the central magnetar in the
observer frame, $M_{\mathrm{sw}}=\left( 4/3\right) \pi r^{3}nm_{p}$ is the
swept-up mass of the ambient medium (region 1), $\gamma \simeq \gamma _{3}$
is the Lorentz factor of the forward-shocked medium (region 2), $\gamma _{%
\mathrm{ej},0}$ is the initial Lorentz factor of the ejecta. In the
analytical calculations, we set $\gamma _{\mathrm{ej},0}=1$, and in the
numerical calculations discussed in Section \ref{sec-PTF11agg} we set $%
\gamma _{\mathrm{ej},0}$ according to the typical initial velocity of the
ejecta $\beta _{\mathrm{ej},0}\simeq 0.2$.\ Because the fraction of the
total power is $\xi $ $\sim 0.8-0.9$ \citep{zhang13}, we will approximate it
as $\xi \approx 1$ in the following calculation. Equation $\left( \ref%
{dynamics}\right) $ is different from equation (1) of \cite{gao13} by a
factor 2 in the second term on the right hand side because the injected
energy is deposited both in FS and in RS and the energy contained in FS and
RS is comparable \citep{blandford76}.

The FS emission is calculated quantitatively similar to that derived by \cite%
{gao13}. The energy density and number density of the reverse-shocked wind
(region 3) is determined by \citep{sari95,blandford76} $%
e_{3}/(n_{3}m_{e}c^{2})=\bar{\gamma}_{3}-1\simeq \bar{\gamma}_{3}$ and $%
n_{3}/n_{4}=4\bar{\gamma}_{3}+3\simeq 4\bar{\gamma}_{3}$ with $%
n_{4}=L_{0}/4\pi r^{2}\gamma _{4}^{2}m_{e}c^{3}$, where $\gamma _{4}$ is the
Lorentz factor of the unshocked wind (region 4). The minimum Lorentz factor
of the $e^{+}e^{-}$ in region 3 is \citep{sari98} $\gamma _{3m}=\epsilon
_{e}[(p-2)/\left( p-1\right) ]\bar{\gamma}_{3}$, where a constant fraction $%
\epsilon _{e}$ (subject to the condition $\epsilon _{e}+\epsilon _{B}=1$) of
the shock energy goes into $e^{+}e^{-}$ so that the magnetic field of region
3 is determined by $B_{3}=\left( 8\pi \epsilon _{B}e_{3}\right) ^{1/2}$ %
\citep{sari98}. The self-absorption frequency $\nu _{a}$ is calculated
according to \cite{wu03}.

Before the FS becomes relativistic, the slow expansion of the ejecta implies
a large $\bar{\gamma}_{3}\simeq \gamma _{4}$. Therefore $e^{+}e^{-}$ in
region 3 are very hot, resulting in high X-ray flux, which will last for
several hundred seconds before optical emission takes over. The continuous
energy injection will maintain the optical flux to a relatively high level
until the time $T_{\mathrm{sd}}$. During this process, radio emission
becomes progressively dominated and could last for years before rapid
decline. At time $T_{\mathrm{sd}}$, the central engine turns off and region
4 disappears, thereafter region 3 begins to spread linearly, i.e., the width
of region 3 in comoving frame $\Delta _{3}\propto r$. Additionally, because
the $e^{+}e^{-}$ are in slow cooling regime after $T_{\mathrm{sd}}$ (see
Figure \ref{fig-nu}), $\gamma _{3m}$ is essentially constant thereafter.

\cite{gao13} discussed four cases depending on different parameter
configurations. Here we consider only Case I in their paper, viz. the case $%
T_{\mathrm{dec}}<T_{\mathrm{sd}}$. We find that the optical transient
PTF11agg \citep{cenko13}\ can be neatly interpreted according to RS emission
powered by a millisecond magnetar (Figure \ref{fig-flux}).

\begin{figure}[tbph]
\centering\includegraphics[height=13.cm,angle=-90]{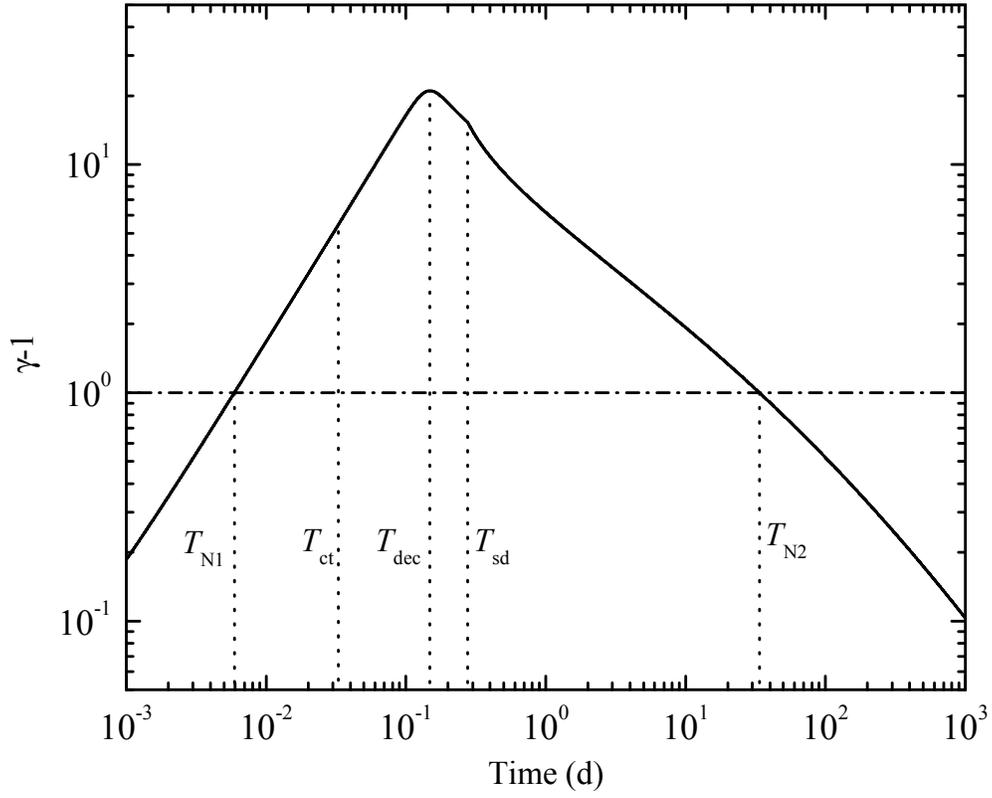}
\caption{Evolution of the ejecta's Lorentz factor of PTF11agg.}
\label{fig-gamma}
\end{figure}

\begin{figure}[tbph]
\centering\includegraphics[height=13.cm,angle=-90]{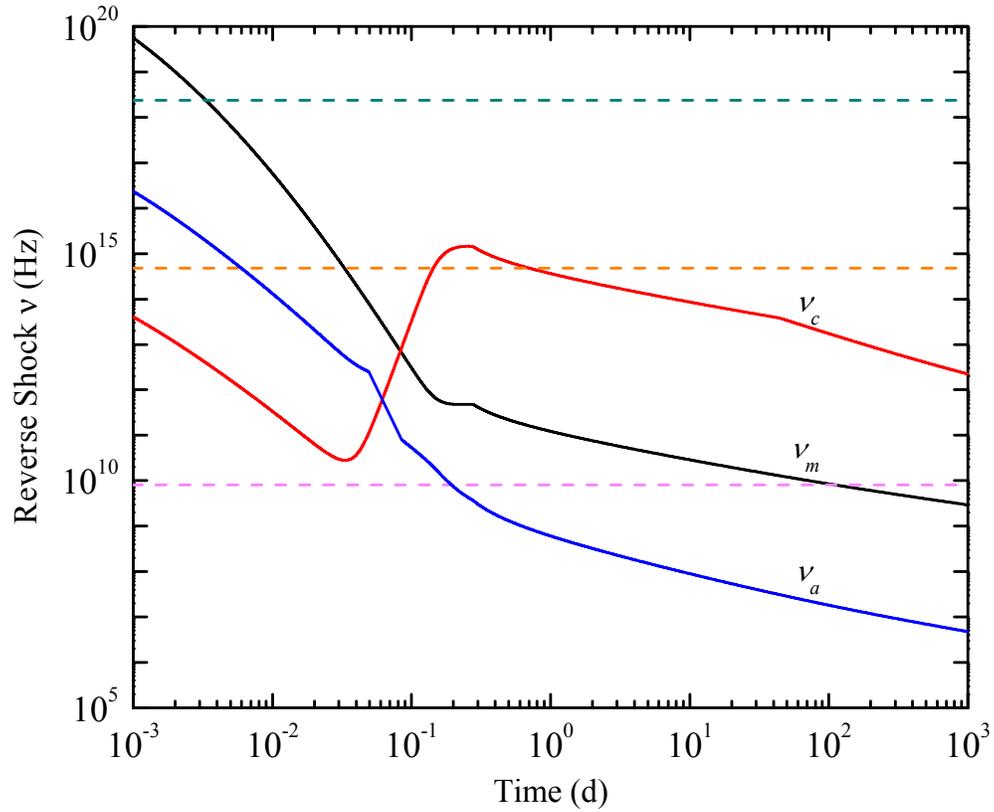}
\caption{Characteristic frequencies of PTF11agg (Numerical results). The
three dashed lines mark radio (8 $\unit{GHz}$), optical ($R$) and X-ray
bands, respectively.}
\label{fig-nu}
\end{figure}

Here we first list the corresponding timescales and Lorentz factor at $T_{%
\mathrm{dec}}$ that are similar to that derived by \cite{gao13}:%
\begin{eqnarray}
T_{\mathrm{N1}} &=&2.1\times 10^{-2}\unit{d}L_{0,47}^{-1}M_{\mathrm{ej},-4}
\\
T_{\mathrm{dec}} &=&0.28\unit{d}L_{0,47}^{-7/10}M_{\mathrm{ej}%
,-4}^{4/5}n^{-1/10} \\
T_{\mathrm{N2}} &=&2.4\times 10^{2}\unit{d}L_{0,47}^{1/3}T_{\mathrm{sd}%
,5}^{1/3}n^{-1/3} \\
\gamma _{\mathrm{dec}} &=&6.7L_{0,47}^{3/10}M_{\mathrm{ej}%
,-4}^{-1/5}n^{-1/10}+1,
\end{eqnarray}%
where we use day instead of second as the units of time to ease comparison
with the observational data (Figure \ref{fig-gamma} and \ref{fig-flux}).
Here $T_{\mathrm{N1}}$ and $T_{\mathrm{N2}}$ are the times $\gamma -1=1$,
viz. the transition times between relativistic and Newtonian dynamics, $T_{%
\mathrm{dec}}$ is the deceleration time whereafter the blast wave begins to
decelerate. It can be seen that the above analytically derived values are in
good agreement with the numerical results (Figure \ref{fig-gamma}) except
for $T_{\mathrm{N2}}$, which could be overestimated by one magnitude. The
reason is that we set the radius $r_{\mathrm{N2}}=cT_{\mathrm{N2}}$ in the
analytical calculations. This approximation significantly underestimates the
actual radius because of the prominent relativistic time propagation effect %
\citep{zhang04} before $T_{\mathrm{N2}}$. A better approximation is to take $%
r_{\mathrm{N2}}=4\bar{\gamma}^{2}cT_{\mathrm{N2}}$ with the average Lorentz
factor $\bar{\gamma}$ lying between $1$ and $\gamma _{\mathrm{dec}}$.

The temporal scaling indices of various parameters are listed in Table \ref%
{tbl-indices}. From Figure \ref{fig-nu} and Table \ref{tbl-indices} we see
that there are three more times that shape the temporal evolution of the
corresponding parameters and lightcurves:
\begin{eqnarray}
T_{\mathrm{ct}} &=&6.0\times 10^{-2}\unit{d}L_{0,47}^{-2/3}M_{\mathrm{ej}%
,-4}^{5/6}\epsilon _{B,-1}^{1/6} \\
T_{ac} &=&0.11\unit{d}L_{0,47}^{-\left( 8p+25\right) /2\left( 6p+19\right)
}M_{\mathrm{ej},-4}^{\left( 5p+16\right) /\left( 6p+19\right) }\epsilon
_{B,-1}^{\left( 2p+5\right) /2\left( 6p+19\right) }\gamma _{4,4}^{-1/\left(
6p+19\right) } \\
T_{mc} &=&0.13\unit{d}L_{0,47}^{-5/7}M_{\mathrm{ej},-4}^{6/7}\epsilon
_{B,-1}^{1/7}\epsilon _{e}^{1/7}\gamma _{4,4}^{1/7},
\end{eqnarray}%
where $T_{ac}$ and $T_{mc}$ are the respective crossing time of $\nu _{c}$
with $\nu _{a}$ and $\nu _{m}$. More words are needed for $T_{\mathrm{ct}}$.
Owing to the brake caused by the massive ejecta, the magnetar wind cannot
drive the ejecta to large radius at beginning, resulting in high energy
density and therefore strong magnetic field of region 3. Consequently, the $%
e^{+}e^{-}$ cool so fast that the cooling Lorentz factor \citep{sari98} $%
\gamma _{c}\approx 1$. Only after $T_{\mathrm{ct}}$ does $\gamma _{c}$
deviate significantly from 1. So $T_{\mathrm{ct}}$ is defined by the
condition $3\pi m_{e}c=\sigma _{T}\gamma B_{3}^{2}T_{\mathrm{ct}}$.

The various frequencies of the RS emission at $T_{\mathrm{dec}}$ are
\begin{eqnarray}
\nu _{a,\mathrm{dec}} &=&1.8\times 10^{10}\unit{Hz}L_{0,47}^{51/50}M_{%
\mathrm{ej},-4}^{-12/25}\epsilon _{B,-1}^{1/5}\epsilon_{e}^{-1}\gamma
_{4,4}^{-8/5}n^{13/50}\frac{p-1}{p-2}\left[ \frac{p+2}{p+\frac{2}{3}}\left(
p-1\right) \right] ^{3/5} \\
\nu _{m,\mathrm{dec}} &=&9.9\times 10^{12}\unit{Hz}\epsilon
_{B,-1}^{1/2}\epsilon _{e}^{2}\gamma _{4,4}^{2}n^{1/2}\left( \frac{p-2}{p-1}%
\right) ^{2} \\
\nu _{c,\mathrm{dec}} &=&5.1\times 10^{14}\unit{Hz}L_{0,47}^{1/5}M_{\mathrm{%
ej},-4}^{-4/5}\epsilon_{B,-1}^{-3/2}n^{-9/10} \\
F_{\nu ,\max ,\mathrm{dec}} &=&5.6\times 10^{3}\unit{mJy}L_{0,47}^{9/10}M_{%
\mathrm{ej},-4}^{2/5}\epsilon _{B,-1}^{1/2}\gamma
_{4,4}^{-1}n^{1/5}D_{27}^{-2}.
\end{eqnarray}

\section{Numerical Approach and the Transient Source PTF11agg}

\label{sec-PTF11agg}

Recently, the wide-field survey telescope Palomar Transient Factory (PTF)
reported the discovery of a transient source, PTF11agg, of very unusual
nature \citep{cenko13}. PTF11agg consists of a bright, rapidly fading
optical transient of two days long and an associated year-long scintillating
radio transient, without a high-energy trigger. It is demonstrated that a
galactic origin of such a transient is ruled out \citep{cenko13}. We inspect
the observed properties and lightcurves of PTF11agg and speculate that this
transient could be the first evidence for RS emission powered by a magnetar.

There are several lines of reasoning for this speculation. First, a magnetar
wind can power the optical RS emission until $T_{\mathrm{sd}}$, which is
typically 1 day. Second, the duration of radio emission of PTF11agg is in
accord with our estimate of the duration of radio RS emission powered by a
millisecond magnetar. Third, the energy scale of the blast wave of PTF11agg
is just the same as the rotational energy of a millisecond magnetar. It is
measured by means of interstellar scattering and scintillation that PTF11agg
had an angular diameter of $\Theta \approx 20\unit{\mu as}$ at observer's
time $\Delta t_{\mathrm{obs}}\approx 100$ days. By this time the emitting
source should be in the transrelativistic or Newtonian regime so that the
Sedov-Taylor energy is approximately applicable, $E_{0}=(4/3)\pi
r^{3}nm_{p}c^{2}$. Adopting the typical values $n\simeq 1\unit{cm}^{-3}$ and
$z\simeq 1$, we estimate the total energy injected to be $\sim 10^{52}\unit{%
erg}$, i.e., the energy scale of a typical millisecond magnetar. Fourth,
although the simplest on-axis long GRB (LGRB) afterglow explanation proposed
by \cite{cenko13} cannot be ruled out at this time, we will see (Figure \ref%
{fig-flux}) that the magnetar model proposed in this paper provides a much
better fit to the data with more reasonable fitting parameters.

Based on the above lines of reasoning, we perform numerical calculations as
well as analytical calculations presented above. In our numerical
calculations, we first solve equation $\left( \ref{dynamics}\right) $ for $%
\gamma $, from which the velocity of the ejecta can be got. We then precede
to accumulatively calculate the radius of the shock front. $M_{\mathrm{sw}}$%
, $n_{4}$ and other quantities are then calculated straightforwardly.
Because we do not know a priori the redshift of the source, we just guess a
redshift in the range $0.5\lesssim z\lesssim 3.0$ as constrained by \cite%
{cenko13}. Then we determine a group of parameters that best fit the data
for such a redshift. If the resulting size of the blast wave at $\Delta t_{%
\mathrm{obs}}\approx 100$ days does not satisfactorily give the measured
angular size, we guess another redshift until a self-consistent fitting is
found. In passing, although we do not bother to include the redshift $z$ in
equations listed above, we do include it in the numerical calculations.

The initial radius of the swept region is set to $r_{0}=6\times 10^{10}\unit{%
cm}$ which is the distance the ejecta, with a typical velocity $v\approx
0.2c $, traveled before the magnetar wind is launched \citep{gao13}. The
numerically determined best-fit lightcurves are depicted in Figure \ref%
{fig-flux} with the analytical result marked piecewisely. The
best-fit parameters are listed in Table \ref{tbl-parameter}. In
Figure \ref{fig-flux} we do not include the two points (standing for
a faint quiescent optical counterpart) in the optical lightcurve at
later times \citep[i.e., $t>10\unit{days}$, see Figure 6
of][]{cenko13} because by these times the optical flux coming from
the transient source fell below $0.1\unit{\mu Jy}$ so that the faint
quiescent optical counterpart stood out. We find that
the magnetar wind was launched at $t_{0}=0$:55:06 on 2011 January 30, i.e., $%
4.368$ hours before the first image was taken at 5:17:11 on the same day %
\citep{cenko13}. This launch time is more than 1 hour later than determined
by \cite{cenko13}, resulting in a shallower optical decay index.

From Table \ref{tbl-parameter} we see that $M_{\mathrm{ej}}=1.2\times
10^{-4}M_{\odot }$, consistent with the numerical simulations %
\citep{rezzolla10}. The luminosity $L_{0}$ and the local frame spin
down timescale $T_{\mathrm{sd}}/\left( 1+z\right) $ of the magnetar
give the initial rotation period of the magnetar
$P_{0}=3.1\unit{ms}$ in local frame and the dipole magnetic field
$B_{p}=2.0\times 10^{15}\unit{G}$ for the typical values $R_{6}=1$
and $I_{45}=1.5$. The derived $P_{0}=3.1\unit{ms}$
lies between the minimum rotation period of a stable NS $P_{\mathrm{crit}%
}\sim 1\unit{ms}$ and the maximum rotation period $P_{\mathrm{dyn}}\sim 5%
\unit{ms}$ when the $\alpha $-$\Omega $ dynamo action quickly builds up the
magnetic field before the NS dissipates its internal heat via cooling %
\citep{duncan92,thompson93}. $\gamma _{4}=4.6\times 10^{4}$ is consistent
with the value derived for the Crab pulsar \citep{atoyan99,dai04}. The
inferred circumburst density $n=0.13\unit{cm}^{-3}$ is consistent with the
findings by other studies \citep{berger05,soderberg06,berger07}. In
contrast, the circumburst density for LGRBs is usually higher than for
SGRBs. This strengthens the argument that PTF11agg was a circum-binary
transient rather than a LGRB afterglow considered by \cite{cenko13} because
they inferred $n\sim 10^{-3}\unit{cm}^{-3}$. We therefore conclude that the
best-fit values are all well within the reasonable ranges.

Figure \ref{fig-flux} shows that the temporal decay indices of optical flux
and radio flux both depend on the parameter $p$. This same parameter also
sensitively determines the time $T_{m,\mathrm{rad}}$ when $\nu _{m}=\nu _{%
\mathrm{rad}}$ because $\nu _{m}\propto \left[ \left( p-2\right) /\left(
p-1\right) \right] ^{2}$. In the numerical calculations we find that $p$ is
accurately determined as $p=2.2$, any deviation of even $0.05$ would \textit{%
significantly} modify the resulting lightcurves so as not to fit the data
closely. $T_{\mathrm{sd}}$ is also accurately determined as can be seen from
the optical lightcurve in Figure \ref{fig-flux}. To get a satisfactory fit, $%
n$ cannot deviate from the given value by more than $0.08$. The only loosely
constrained value is $\epsilon _{B}$, for which a deviation of $0.05$ is
also acceptable. But too large a value of $\epsilon _{B}\sim 0.2$ is not
favored by the data.

Figure \ref{fig-flux} shows that the radio flux suffered from a rapid
decline, with a temporal decay index $\alpha =3\left( p+1\right) /10\simeq 1$%
, after the time $T_{m,\mathrm{rad}}\approx 113$ days. Before $T_{m,\mathrm{%
rad}}$ the derived radio spectral index $\beta =1/3$ is consistent with the
observations \citep[see
Figure 4 of][]{cenko13}. We infer $\gamma =1.9$ at $\Delta t_{\mathrm{obs}%
}\approx 42$ days and $\gamma =1.5$ at $\Delta t_{\mathrm{obs}}\approx 100$
days, consistent with the limits derived by \cite{cenko13}. In fact, the
ejecta gained a maximum Lorentz factor $\gamma \simeq 22$ at $T_{\mathrm{dec}%
}$ (see Figure \ref{fig-gamma}). To get a high Lorentz factor, as is
required by PTF11agg, in our model that is baryon polluted, the parameters
should be tuned so that $T_{\mathrm{dec}}\simeq T_{\mathrm{sd}}$, in which
case $\gamma $ can be as large as $\gtrsim 50$ for the typical values we
adopt \citep[i.e. Case II discussed
by][]{gao13}. This is nearly the case for PTF11agg because we find $T_{%
\mathrm{dec}}=0.15\unit{d}$ and $T_{\mathrm{sd}}=0.28\unit{d}$, i.e.,
PTF11agg in fact lies between Case I and Case II discussed by \cite{gao13}.

Of particular interest is the optical lightcurve between $T_{\mathrm{dec}}$
and $T_{\mathrm{sd}}$ (see Figure \ref{fig-flux}). Analytical calculation
shows that the flux density in this time interval $F_{\nu ,\mathrm{opt}%
}\propto F_{\nu ,\max }\nu _{m}^{\left( p-1\right) /2}$ with $F_{\nu ,\max
}\propto t^{1/2}$ and $\nu _{m}$ first declines and then flattens (see Table %
\ref{tbl-indices}). This behavior nicely accounts for the observed optical
lightcurve. We mention that, as seen from Figure \ref{fig-flux}, the FS
emission is negligible compared to the RS emission.

The total injected energy $E_{0}=L_{0}T_{\mathrm{sd}}$ in the observer frame
implies a total fluence of $S_{\mathrm{bol}}=2.4\times 10^{-7}\unit{erg}%
\unit{cm}^{-2}$, which is well below the $\gamma $-ray sensitivity to
fluences $\left( 10\unit{keV}-5\unit{MeV}\right) $ of $S_{\gamma }\gtrsim
6\times 10^{-7}\unit{erg}\unit{cm}^{-2}$ of the Third InterPlanetary Network
(IPN) with essentially all-sky coverage \citep{cenko13}. Other detectors
with a higher sensitivity cover only a narrow field of view, e.g., $8.8\unit{%
sr}$ and $2\unit{sr}$ for GBM and \textit{Swift} BAT respectively, therefore
missing the very early high-energy emission with a high probability.

We also try to fit the lightcurves without RS involved and find that no good
fit can be achieved under the simple assumptions such as constant $\epsilon
_{e}$ and $\epsilon _{B}$ and $2<p<3$.

\section{Discussion and Conclusions}

\label{sec-discuss}

In this paper we suggest that at least a fraction of BNS mergers produce
massive NSs rather than black holes. The ensuing dynamo actions operate to
boost the magnetic field to the magnetar level. The rotational energy of the
central magnetars is injected into the ejecta as Poynting flux, which could
become lepton dominated so that strong RS could be developed. The optical RS
emission could last for $\sim 1$ day and radio emission for years. We then
apply our model to the optical transient PTF11agg.

\begin{figure}[tbph]
\centering\includegraphics[height=13.cm,angle=90]{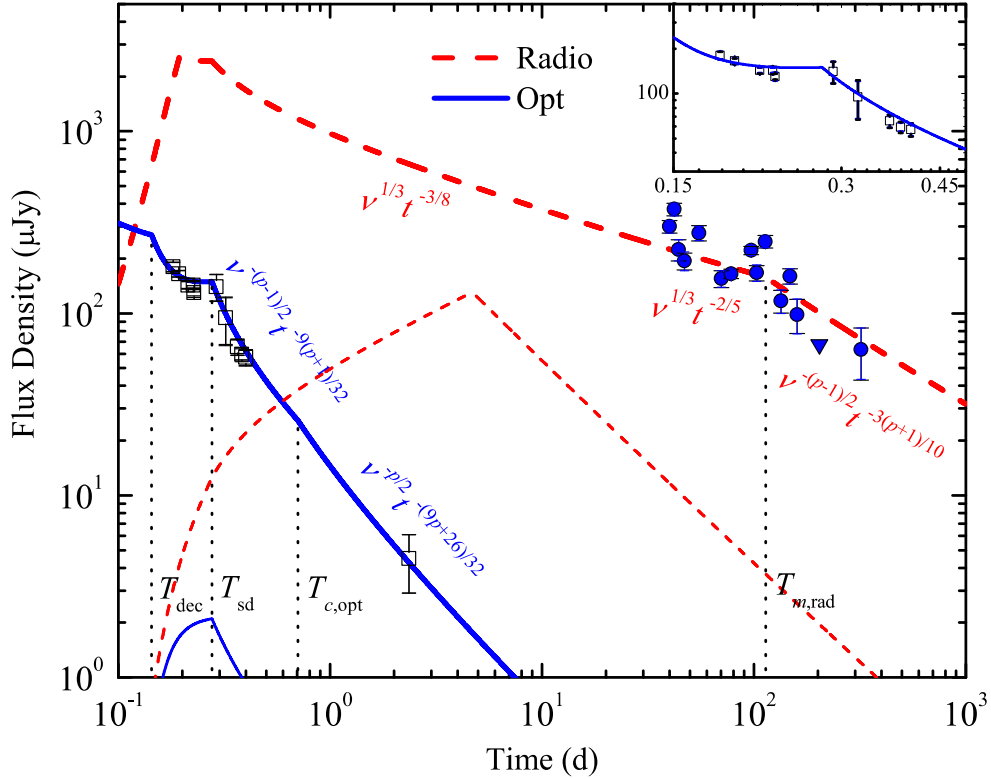}
\caption{Fitting to the observed data. Solid lines correspond to optical
light curves, and dashed lines to radio light curves, among which thick
curves are for RS emission and thin curves for FS emission. Dotted vertical
lines indicate the characteristic times. $T_{c,\mathrm{opt}}$ is the time $%
\protect\nu _{c}=\protect\nu _{\mathrm{opt}}$ whereafter the cooling
frequency $\protect\nu _{c}$ falls below the optical bandpass $\protect\nu _{%
\mathrm{opt}}$, $T_{m,\mathrm{rad}}$ is the time $\protect\nu _{m}=\protect%
\nu _{\mathrm{rad}}$ whereafter the minimum frequency $\protect\nu
_{m}$ falls below the radio bandpass $\protect\nu _{\mathrm{rad}}$.
Inset is the zoom-in of the optical light curve. The data points are
taken from \cite{cenko13}.} \label{fig-flux}
\end{figure}

To interpret the observed lightcurves of PTF11agg, three possibilities were
considered by \cite{cenko13}: an untriggered LGRB, an orphan afterglow due
to viewing-angle effects, and a dirty or failed fireball. The untriggered
GRB interpretation is not favored because the a posteriori detection
probability is only $2.6\%$ in the high-cadence field where PTF11agg was
detected \citep{cenko13}. The orphan afterglow interpretation is also
marginal because it requires that the observer's sightline cannot be outside
the jet opening angle \citep{cenko13}. While the third explanation %
\citep{dermer00,huang02,rhoads03} is possible, the fit to the data is not as
good as that in Figure \ref{fig-flux}, as far as we know.

Consequently, we suggest that PTF11agg may represent the first evidence for
the RS emission powered by a post-merger millisecond magnetar. The magnetars
formed by other scenarios, such as supernova collapse, cannot be the
candidates for PTF11agg because in these scenarios $M_{\mathrm{ej}}\gtrsim
10M_{\odot }$ and the ejecta can never reach a relativistic speed.

Comparison of Figure \ref{fig-flux} with Figure 6 of \cite{cenko13} shows
that the predicted radio lightcurves are quite different, especially in the
early time duration. Consequently, to differentiate this explanation from
the LGRB afterglow model, early observations of the radio lightcurve are
crucial. Another differentiation is the gravitational wave (GW) associated
with the preceding NS-NS merger. The next generation of GW detectors %
\citep{acernese08,abbott09,kuroda10} are promising in detecting GW signals
from nearby PTF11agg-like compact binary mergers up to a distance $\lesssim
100\unit{Mpc}$. Other EM signals, including SGRBs, radio afterglows, optical
macronovae, and X-ray emissions are also helpful in identifying post-merger
magnetars. To confirm the binary-merger nature of a source like PTF11agg at
cosmological distances, i.e., $z\gtrsim 1$, however, the most promising
counterpart is SGRBs. But one should be aware of the caveat that SGRBs can
only be observed in a narrow angle and the RS emission discussed in our
model is weakest in this angle \citep[see
Figure 1 of][]{gao13}.

\begin{acknowledgements}
We thank the anonymous referee for valuable comments and
constructive suggestions, and He Gao, Xuan Ding, and Xue-Feng Wu for
helpful discussions. This work is supported by the National Natural
Science Foundation of China (grant No. 11033002).
\end{acknowledgements}

\clearpage
\begin{table}[tbp]
\caption{Analytical Temporal Scaling Indices of Different Parameters of the
RS.}
\label{tbl-indices}
\begin{center}
\begin{tabular}{lcccccc}
\hline\hline
& $\gamma -1$ & $r$ & $\nu _{a}$ & $\nu _{m}$ & $\nu _{c}$ & $F_{\nu ,\max }$
\\ \hline
$t<T_{N1}$ & $1$ & $\frac{3}{2}$ & $-\frac{3p+14}{2\left( p+4\right) }$ & $-%
\frac{3}{2}$ & $-\frac{3}{2}$ & $-\frac{1}{2}$ \\
$T_{N1}<t<T_{\mathrm{ct}}$ & $1$ & $3$ & $-\frac{3p+14}{p+4}$ & $-5$ & $-3$
& $-2$ \\
$T_{\mathrm{ct}}<t<T_{ac}$ & $1$ & $3$ & $-\frac{3p+2}{p+4}$ & $-5$ & $9$ & $%
-2$ \\
$T_{ac}<t<T_{mc}$ & $1$ & $3$ & $-\frac{6\left( 3p+2\right) }{5}$ & $-5$ & $%
9 $ & $-2$ \\
$T_{mc}<t<T_{\mathrm{dec}}$ & $1$ & $3$ & $-\frac{13}{5}$ & $-5$ & $9$ & $-2$
\\
$T_{\mathrm{dec}}<t<T_{\mathrm{sd}}$ & $-\frac{1}{4}$ & $\frac{1}{2}$ & $-%
\frac{3}{5}$ & $0$ & $-1$ & $\frac{1}{2}$ \\
$T_{\mathrm{sd}}<t<T_{\mathrm{N2}}$ & $-\frac{3}{8}$ & $\frac{1}{4}$ & $-%
\frac{3}{4}$ & $-\frac{9}{16}$ & $-\frac{17}{16}$ & $-\frac{9}{16}$ \\
$T_{\mathrm{N2}}<t$ & $-\frac{6}{5}$ & $\frac{2}{5}$ & $-\frac{18}{25}$ & $-%
\frac{3}{5}$ & $-\frac{3}{5}$ & $-\frac{3}{5}$ \\ \hline
\end{tabular}%
\end{center}
\end{table}
\begin{table}[tbp]
\caption{Best-fit RS parameters for PTF11agg.}
\label{tbl-parameter}
\begin{center}
\begin{tabular}{cccccccc}
\hline\hline
$L_{0,47}$ & $T_{\mathrm{sd}}$ & $M_{\mathrm{ej},-4}$ & $n$ & $\gamma _{4,4}$
& $p$ & $\epsilon _{B}$ & $z$ \\
$\unit{erg}\unit{s}^{-1}$ & $\unit{d}$ & $M_{\odot }$ & $\unit{cm}^{-3}$ &
&  &  &  \\ \hline
$4.1$ & $0.28$ & $1.2$ & $0.13$ & $4.6$ & $2.2$ & $0.1$ & $2.2$ \\ \hline
\end{tabular}%
\end{center}
\end{table}

\clearpage

\end{document}